\documentclass[aps,prx,twocolumn,final,letterpaper]{revtex4}

\usepackage{appendix}
\usepackage{graphicx}   
\usepackage{import}                         
\usepackage{epstopdf}
\usepackage{amsmath} 
\usepackage{bm}
\usepackage{amssymb}
\usepackage{quotes}
\usepackage{transparent}
\usepackage{dcolumn}
\usepackage{multirow}
\usepackage{cancel} 
\usepackage{mdframed}
\usepackage{color}
\usepackage{bm}
\usepackage{dsfont}
\usepackage{slashed}
\usepackage{enumitem}

\begin{document}
\rmfamily
\raggedbottom

%-----TITLE-----
\title{Noise-immune quantum correlations of intense light}

%-----AUTHORS AND AFFILIATIONS-----
\author{Shiekh Zia Uddin$^{1,2,\dagger}$, Nicholas Rivera$^{1,3,4\dagger}$, Devin Seyler$^{2}$, Jamison Sloan$^{2}$, Yannick Salamin$^{1,2}$, Charles Roques-Carmes$^{2,5}$, Shutao Xu$^{6}$, Michelle Y. Sander$^{6,7,8}$, Ido Kaminer$^{9}$, and Marin Solja\v{c}i\'{c}$^{1,2}$}
\affiliation{$\dagger$ These authors contributed equally. E-mail: nrivera@fas.harvard.edu, suddin@mit.edu
\\
$^{1}$Department of Physics, Massachusetts Institute of Technology, Cambridge, MA 02139, USA. 
\\
$^{2}$Research Laboratory of Electronics, Massachusetts Institute of Technology, Cambridge, MA 02139, USA.
\\
$^{3}$Department of Physics, Harvard University, Cambridge, MA 02138, USA.
\\
$^{4}$School of Applied and Engineering Physics, Cornell University, Ithaca, NY 14853, USA.
\\
$^5$ E. L. Ginzton Laboratory, Stanford University, Stanford, California 94305, USA.
\\
$^6$Department of Electrical and Computer Engineering and BU Photonics Center, Boston University, Boston, MA 02215, USA.
\\
$^7$Division of Materials Science and Engineering, Boston University, Brookline, MA 02446, USA.
\\
$^8$Department of Biomedical Engineering, Boston University, Boston, MA 02215, USA.
\\
$^9$Department of Electrical and Computer Engineering, Technion-Israel Institute of Technology, Haifa, Israel.}

\begin{abstract} 
Lasers with high intensity generally exhibit strong intensity fluctuations far above the shot-noise level. Taming this noise is pivotal to a wide range of applications, both classical and quantum. Here, we demonstrate the creation of intense light with quantum levels of noise even when starting from inputs with large amounts of excess noise. In particular, we demonstrate how intense squeezed light with intensities approaching 0.1 $\text{TW/cm}^2$, but noise at or below the shot noise level, can be produced from noisy inputs associated with high-power amplified laser sources (an overall noise-reduction of 30-fold). Based on a new theory of quantum noise in multimode systems, we show that the ability to generate quantum light from noisy inputs results from multimode quantum correlations, which maximally decouple the output light from the dominant noise channels in the input light. As an example, we demonstrate this effect for femtosecond pulses in nonlinear fibers, but the noise-immune correlations that enable our results are generic to many other nonlinear systems in optics and beyond. 
\end{abstract}

\maketitle

Lasers are among the most important inventions of the twentieth century, with innovations continuing more than sixty years after the demonstration of the first laser \cite{maiman1960stimulated}. While lasers are much more stable than the thermal light sources preceding them, their intensity and phase still fluctuate, due to both external sources and intrinsic quantum-mechanical effects. In many settings including metrology \cite{ligo2011gravitational,taylor2013biological,casacio2021quantum}, quantum state preparation \cite{walls1983squeezed, kaufman2021quantum}, and biomedical imaging \cite{taylor2013biological,casacio2021quantum, rao2021shot}, it is important to have both high intensity and low intensity fluctuations, e.g., at, or even below, the shot-noise level associated with the quantum-mechanical coherent states of Glauber. For shot noise (equivalently, Poissonian statistics), the variance $(\Delta n)^2$ in the detected photon number $n$ is given by $(\Delta n)^2 = \langle n \rangle$, with $\langle n \rangle$ the mean photon number. 

For lower-intensity lasers, such as continuous-wave sources based on stable optical oscillators, the intensity noise often follows the expectation from Poisson statistics. In contrast, higher-power laser systems – which are typically based on amplification of a lower-intensity laser \cite{mourou1998ultrahigh,mourou2006optics} as shown in Fig. 1a – tend to deviate from Poissonian statistics, frequently displaying intensity fluctuations orders-of-magnitude in excess of the shot-noise limit. This excess noise often comes from the amplifier, where it is known that for a linear phase-insensitive amplifier, the intensity variance scales as the square of the power gain \cite{loudon2000quantum, caves1982quantum}. Various techniques exist to control noise, such as linear filtering \cite{haus2000electromagnetic}, active feedback \cite{bachor2019guide}, or passive techniques based on second- or third-order optical nonlinearities \cite{walls1983squeezed, slusher1985observation, bergman1991squeezing, andersen201630}. However, these techniques are often unable to achieve sufficient noise reduction, particularly when the initial noise level is high, without extreme attenuation. As a result, it is difficult to produce light with high intensity and quantum levels of noise. Addressing this roadblock could enable, beyond the applications described above, a wide range of new techniques for generating quantum states of light \cite{spasibko2017multiphoton,heimerl2024multiphoton,gorlach2023high}, as well as quantum states of excitations in materials such as electrons \cite{mitrano2016possible,kennes2017transient} and phonons \cite{garrett1997vacuum,knap2016dynamical,cartella2018parametric,disa2021engineering} – based on strong-field driving, where the needed intensities are achievable, but the requisite noise levels are yet out of reach. 

\begin{figure*}[t]
     \centering
     \includegraphics[width=0.8\linewidth]{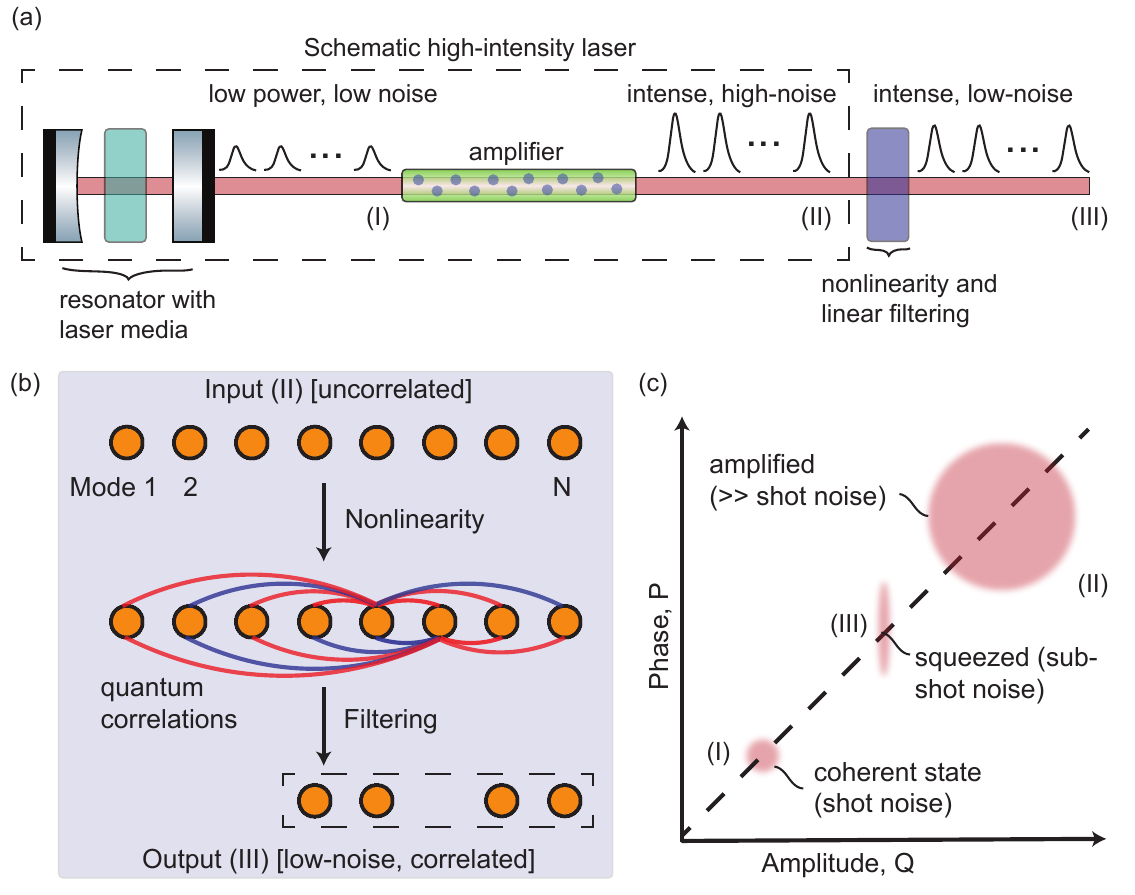}
     \caption{\textbf{High-intensity sources of light with quantum levels of noise.}  (a) A typical high-intensity laser system is realized by amplifying the output of a lower intensity laser which can generate approximations of coherent states.  After amplification, the intensity and noise are both strongly enhanced. By subjecting the output of a high-intensity source to nonlinearity, such as third-order nonlinearity plus mode-selective filtering, one can lower the noise considerably with modest attenuation – in some cases producing quantum light, such as squeezed states, from noisy inputs. (b) The principle for noise reduction is that nonlinearity creates quantum intensity correlations between different modes (yellow circles) in the laser field (red line $-$ positive correlation, blue line $-$ negative). Filtering through a subset of those modes can lead to an output whose correlations “conspire” to make the remaining modes immune to noise at the input. (c) The noise at various stages in (a) is shown in terms of phase space distributions. In this work, we demonstrate the step from (II) to (III).
    }
     \label{fig:fig1}
 \end{figure*}

Our work is based on a surprising discovery that intense pulses propagating through a nonlinear medium can emerge with fluctuations at or below the shot-noise level (in a squeezed state), despite having initial fluctuations far above the shot-noise limit. This is realized with much lower attenuation than when employing pure linear attenuation for noise reduction. We demonstrate this effect and uncover its origin, as a carefully tuned combination of nonlinear dynamics and linear attenuation. Understanding its origin enables us to maximize the squeezing and intensity simultaneously, reaching noise reduction conditions that exceed by nearly an order-of-magnitude the noise reduction expected from pure linear attenuation of the same magnitude. This approach let us create pulses with focusable intensities on the order of 0.1 TW/cm$^2$, with measured noise 4 decibels (dB) below the shot noise level. 

To capture the entire picture, we build a new theory of quantum noise in multimode nonlinear systems, with which we show that the nonlinear dissipation process producing the squeezed light is essentially immune to the addition of noise in the pump. Consequently, the output noise is largely independent of the input level of noise. This noise-immune squeezing is a collective effect, arising from correlations between different modes created by nonlinearity: while any individual mode has strong noise, a group of modes, together, can almost completely decouple from noise, even in the noisiest channels of the input. We call these noise-immune quantum correlations. Our results $-$ demonstrated for the example of femtosecond pulses propagating through an optical fiber $-$ are general, applying to a wide range of nonlinearities that create quantum correlations between different modes.

We start by illustrating the core concept of this work: that although high-intensity lasers generally have noise exceeding the shot-noise limit of coherent states, nonlinear effects can lead to noise reduction below the shot noise limit while retaining substantial intensity. High-intensity laser systems are typically realized by amplifying the output of a lower-intensity laser. While the output is also intense, the intensity fluctuations are much larger than the shot-noise level (see Figs. 1a,c). In this work, we will consider what happens when the output of an amplified high-intensity laser goes through a nonlinear filtering or nonlinear dissipation process, which we define as any process which is both nonlinear in the complex amplitudes of the input, \emph{and} changes (reduces) the intensity of the output relative to the input. The intensity must change: otherwise, the intensity fluctuations cannot change. The effect of the nonlinearity will be to induce quantum correlations between different degrees of freedom (modes) in the pulse, see Fig. 1b. Here, we show data for the example where the modes are frequency components of a propagating pulse confined to a single transverse mode. However, the concept also holds when the different degrees of freedom involve spatial modes, polarization, and so on. The filter after the nonlinearity down-selects a subset of the correlated modes which, as we will explain later in the text, are maximally insensitive to noise in the initial pulse. When the insensitivity is sufficiently strong, the noise at the output becomes independent of the noise at the input, permitting the conversion of light far above the shot-noise level into light far below it (corresponding to intensity squeezing). In contrast to previous work exploring squeezing by spectral filtering \cite{friberg1996observation, hirosawa2005photon} of fundamental $N=1$ solitons \cite{rosenbluh1991squeezed,yamamoto1992photon,sizmann1999v, haus2000electromagnetic}, we operate in a regime of much higher power (beyond the regime of fundamental solitons, Fig. 2b) and much higher initial noise levels, allowing us to demonstrate and elucidate the general concept of noise-immune squeezing of intense light, which is the crucial step to high-intensity lasers with quantum levels of noise. 

  \begin{figure*}[t]
     \centering
     \includegraphics[width=0.8\linewidth]{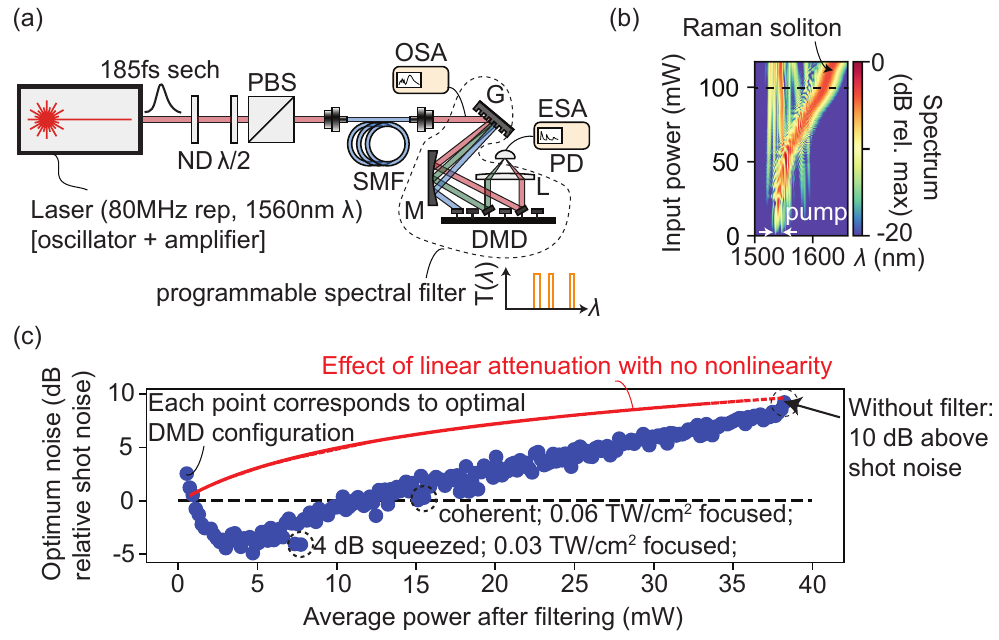}
     \caption{\textbf{Generating intense squeezed light from noisy amplified sources.} (a) Light passes through a single-mode fiber (SMF) and then the output pulse is passed through a programmable spectral filter realized by a grating (G)-mirror (M) pair and a digital micromirror device (DMD). The filtered pulse is sent to a photodiode (PD) and its noise is measured with an electronic spectrum analyzer (ESA). (b) Spectrum of light passing through fiber as a function of average power. Dashed line represents the input power we work at for the remainder of the text. (c) Minimum noise of the light after filtering (blue points), relative to shot noise, as a function of average power transmitted by the filter. The red curve represents the result of frequency-independent linear loss on the output, showing that spectral filtering after nonlinearity leads to much stronger noise reduction and even squeezing, starting from a pulse with noise 10 dB above the shot noise level. The focused intensity is defined as the peak intensity divided by $\pi \lambda^2 / 4$. Legend for additional elements in (a):  filter (ND), half-wave plate ($\lambda/2$), polarizing beam splitter (PBS), lens (L), and optical spectrum analyzer (OSA).  }
     \label{fig:fig2}
 \end{figure*}

We show this effect experimentally. For the amplified laser system, we employ a commercial femtosecond pulsed laser system, which comprises a low-noise oscillator and an erbium-doped fiber amplifier. The resulting pulses are sent through a single-mode fiber (with anomalous dispersion at the center wavelength of the pulses, 1560 nm), which provides nonlinearity. The pulses are then sent into an arbitrarily programmable spectral filter shown in Fig. 2a: the fiber and filter realize the nonlinear dissipation process. In particular, the filter selects a subset of wavelengths which are correlated due to the nonlinearity, acting as a realization of Fig. 1b. Four-wave-mixing and dispersion in the fiber, at high peak pulse intensities, lead to a splitting of the spectrum into a red-shifted soliton (called the Raman soliton), and a highly modulated, chaotic part near the wavelengths of the initial pulse, which we refer to as the pump wavelength (see Fig. 2b). 

Now, we show the effect of filtering after the nonlinearity. Let us consider the space of spectral filters that have the same output power after the filter. For a given output power, there will be an optimal filter that realizes the lowest intensity fluctuations. In general, the noise of the nonlinearly-filtered output is much lower than can be realized by simple broadband linear attenuation alone: at some powers, the difference is approximately an order of magnitude. Moreover, after attenuation, the outputs can be at or below the shot-noise level (up to about 5 dB below the standard quantum limit), despite the input to the nonlinear filtering process having intensity noise which is ten-times the shot-noise limit (see Fig. 2c). Despite attenuation, the peak powers and focusable intensities are high, on the order of 1 kW and 0.1 TW/cm$^2$, respectively.

 \begin{figure}[t]
     \centering
     \includegraphics[width=0.48\textwidth]{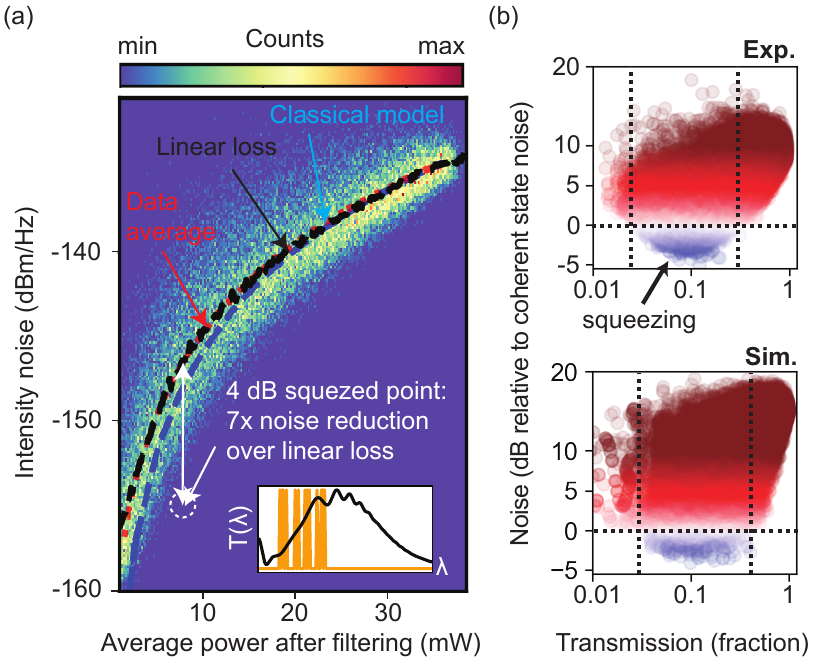}
     \caption{\textbf{Noise statistics of nonlinear filtering.} (a) Filtered intensity (x-axis) and intensity noise (y-axis) for different randomly-chosen filters. Inset represents the type of filter (orange) overlaid on the output spectrum (black; peak corresponds to the red-shifted peak in Fig. 2b) which leads to squeezing. (b, top) Same as in (a) but with the intensity noise normalized to the shot noise level associated with coherent states, as compared to theory (b, bottom). Color scale reflects the noise normalized to shot noise. Dashed lines mark the region of transmission that leads to observed squeezing.}
     \label{fig:fig3}
 \end{figure}
 
Not all filters lead to noise reduction, despite reducing the overall intensity (reducing the intensity by two reduces the intensity noise by two for shot-noise limited light, and four for light with noise far above shot-noise). A randomly chosen filter is about as likely to increase the noise as it is to decrease it: see Fig. 3a, where we plot the output power and intensity noise after a large number of randomly chosen spectral filters (Fig. 3b (top) shows the same result normalized to shot-noise). The minimum noise for each output power corresponds to Fig. 2c. Even for very low attenuation, the noise can decrease or increase by about three decibels. For larger degrees of attenuation, the noise after nonlinear filtering can even increase by an order of magnitude relative to the expectation from linear loss. An example of a filter that leads to strong noise reduction is shown in the inset of Fig. 3a, overlaid on the red-shifted solitonic part of the spectrum in Fig. 2b. The filters that lead to noise reduction overwhelmingly concentrate transmission on the red-side of the solitonic peak. They also tend to feature spectral gaps: without this, noise-reduction is not realized. 

We now describe the physical mechanism underlying the ``noise-immune squeezing'' behavior that we have observed. We start by presenting a general framework we have developed to predict noise in multimode nonlinear systems (even going beyond nonlinear optics). \emph{The theory lets us predict quantum noise purely from classical modeling of the nonlinear dynamics, bypassing second quantization altogether, while nevertheless accurately predicting quantum effects such as squeezing, entanglement, and so on.}. 

Applied to our experimental results: by calculating the classical dynamics of femtosecond pulses in a fiber, according to the generalized nonlinear Schrodinger equation (GNLSE; see Supplementary Information (SI) equation S14) $-$ as is done in all classical work on pulse propagation in fibers $-$ and calculating the derivative of the filtered output intensity with respect to all possible changes in the initial conditions, one can predict quantum noise, bypassing standard approaches such as linearized Heisenberg equations, or stochastic equations. Among the major advantages of our framework, which we call \emph{quantum sensitivity analysis}, are: (1) the resulting noise expressions are directly understood in terms of the classical dynamics, leading directly to new insights, and (2) the framework efficiently and analytically predicts output noise for \emph{arbitrary input noise profiles}, which is challenging for the other methods but necessary for developing intense quantum light. 

While a detailed derivation of the most general form of the theory is provided in the SI, we present a key special case here, connecting the variance of any property of the output to derivatives of the classically predicted output with respect to the initial conditions.  A wide variety of classical nonlinear dynamical systems can be mathematically specified as an initial-value problem, transforming a set of inputs to a set of outputs. Denote the inputs $\alpha_{\text{IN},i}$ and the outputs $\alpha_{\text{OUT},i}$, where $i$ labels a degree of freedom (or mode) and $\alpha$ denotes the value of that degree of freedom. For the system described in Figs. 2 and 3, where the classical dynamics are governed by the GNLSE, the modes $i$ correspond to frequency components of the initial pulse, while the $\alpha$ are the corresponding complex Fourier coefficients. In a general multimode system, the $i$ can also include spatial modes, polarization, and even material degrees of freedom. Any property of the output we measure, denoted $X_{\text{OUT}}$ (e.g., total intensity of a filtered output pulse) is a function of the inputs:  $X_{\text{OUT}} = X_{\text{OUT}}[\{\alpha_{\text{IN}},\alpha^*_{\text{IN}}\}]$ \footnote{Here, we include the dependence on the complex conjugate fields, as they are in principle independent variables, being coupled by nonlinearity.}. 

Assume that the input has noise in the different input modes, denoted $\delta\alpha_{\text{IN},i}$. In the absence of external noise sources, this noise comes only from vacuum fluctuations, described by noise correlations $\langle \delta\alpha_{\text{IN},i}\delta\alpha^*_{\text{IN},j}\rangle = \delta_{ij}, \langle \delta\alpha^*_{\text{IN},i}\delta\alpha_{\text{IN},j}\rangle = \langle \delta\alpha_{\text{IN},i}\delta\alpha_{\text{IN},j}\rangle = \langle \delta\alpha^*_{\text{IN},i}\delta\alpha^*_{\text{IN},j}\rangle = 0$. In the presence of additional noise, from amplifiers or external sources, the noise changes. Assuming that this noise is uncorrelated and phase-insensitive \footnote{This expression can be generalized to include correlated noise, but it is unnecessary to account for our findings.}, as is common in linear amplifiers, the variance in $X_{\text{OUT}}$, denoted $(\Delta X_{\text{OUT}})^2$, is given as (see SI):
\begin{equation}
    (\Delta X_{\text{OUT}})^2 = \sum\limits_i  F_i \Big|\frac{\partial X_{\text{OUT}}}{\partial\alpha_{\text{IN},i}}\Big|_{\boldsymbol{\alpha}_{\text{IN}} = \langle\boldsymbol{\alpha}_{\text{IN}}\rangle, \boldsymbol{\alpha}^*_{\text{IN}} = \langle\boldsymbol{\alpha}_{\text{IN}}\rangle^* }^2.
\end{equation}
Here, the derivative $\partial X_{\text{OUT}}/\partial\alpha_{\text{IN},i}$ represents the sensitivity of the output quantity to changes in the initial conditions of the classical nonlinear dynamics (i.e., it is a purely classical quantity). Additionally, the quantity $F_i$ is the so-called Fano factor which measures the deviation of input mode $i$ from coherent-state statistics. Specifically, it is given by $F_i = (\Delta n_i)^2/\langle n_i\rangle$ $-$ where  $\langle n_i\rangle$ is the mean photon number of the $i$th mode and $(\Delta n_i)^2$ the variance. Importantly, even for vacuum states $\langle n_i\rangle = 0$, $F_i = 1$ and there is a finite and non-negligible contribution to the noise, representing the transduction of quantum vacuum fluctuations of the input modes to the noise in the output quantity. We note that the sum over $i$ is a generalized sum, which for systems with continuous degrees of freedom (such as pulses) is understood as an integral.

In Fig. 3b (bottom), we used this framework to predict the random filtering experiment of Fig. 3a and Fig. 3b (top). The prediction is based on the framework described above. The overall trends, maximum squeezing, and noise-robustness are in excellent agreement. As we show in the Supplementary Information (SI), in Fig. S5, our simulations also predict the same magnitude of squeezing is realized even when the same input (i.e., same spectrum, same average power) has 25 dB more noise than the shot-noise limit (over 300 times), at about the same level of attenuation, indicating very strong noise-immunity \footnote{Importantly, the added noise cannot be removed trivially from the pump (for example by filtering, as is often done in the presence of amplified spontaneous emission), since the added noise is at the wavelengths responsible for the nonlinear dynamics (the pump wavelengths).}. 

Using this theory, we can now elucidate the mechanism of noise-immunity and squeezing. According to Eq. (1), the noise in general increases when the excess noise in some input increases. Moreover, if the excess noise exists in multiple modes, that tends to enhance the noise as well. The only exception, is if $|\partial X_{\text{OUT}}/\partial\alpha_{\text{IN},i}| = 0$ over the range of input modes $i$ where $F_i > 1$. This unusual broadband noise-immunity requires looking at a property which is a combination of different modes (sum of intensities of different Fourier components with different wavelengths, e.g., $X_{\text{OUT}} = \sum_{\lambda} n_{\lambda} $). In this case, fluctuations in the inputs lead to opposite shifts in different output modes, which cancel upon summation, leading to noise immunity, as illustrated in Fig. 4. This allows one to generate squeezing even from highly noisy inputs. Physically, in the case of femtosecond pulses in fibers, this happens because four-wave mixing mixes wavelengths in a photon-number conserving manner, which means that fluctuations (vacuum or otherwise) in the input wavelengths will cause the output intensity at some colors to increase at the expense of others. From this argument, it is clear that \emph{any} conservative mode-mixing nonlinearity can enable this phenomenon, for example: spatially multimode nonlinear systems such as multimode fibers and waveguides.

To make this idea more concrete, we show experimentally that even combining \emph{two} wavelengths, the collective noise can be about an order of magnitude lower than the individual wavelength noises, manifesting this noise immunity due to correlations. In Fig. 5, we consider the example of the sum of photon numbers of two colors (of wavelengths $\lambda,\lambda'$) in the output, denoted $n_{\text{OUT}} = n_{\lambda}+n_{\lambda'}$ (which is realized by using the programmable filter to send two narrow bands of wavelengths to the detector). The noise in $n_{\text{OUT}}$ comes from quantum and other noises in the individual wavelength components of the initial pulse (see Fig. 5a), even those wavelengths which have negligible mean intensity. Certain pairs of wavelengths $\lambda,\lambda'$ at the output display very low sensitivity to noise at wavelengths where the input intensity is concentrated (the input band). This occurs when fluctuations in the input band lead to opposite shifts in the intensities of $\lambda$ and $\lambda'$. For these pairs, the noise is set by vacuum fluctuations in inputs which are initially dark, at wavelengths close to that of the red-shifted soliton (where $F_i = 1$ and the derivative is non-zero, compared to the pump wavelengths where the derivative is zero even though the $F_i \gg 1$). In contrast, other pairs, where fluctuations lead to shifts in the same direction for the two wavelengths, have strong sensitivity to noise in the input band (see Figs. 5b,c). The signature of cancellation is a low noise in the intensity of certain pairs, much lower than the intensity noise of either of the individual wavelengths alone. This effect is consistent with anti-correlated shifts in the two wavelengths in response to noise in many input channels. That noise will not cancel out in the individual wavelength noises, but it will in the ``pair noises''. 

The theoretical predictions for the pair noise and the noise relative to the individual wavelengths, resulting from Eq. (1), are shown in Figs. 5d,e. We measure the pair noises and relative noises, by using our programmable filter to send two wavelengths to the detector, and measuring intensity fluctuations. We scan all possible pairs of wavelengths. The results are shown in Figs. 5f,g, which clearly corroborate the theory, and indicate that at the level of pairs of wavelengths, certain pairs can be an order of magnitude more resilient to noise than the individual wavelengths. 

 \begin{figure}[t]
     \centering
     \includegraphics[width=0.48\textwidth]{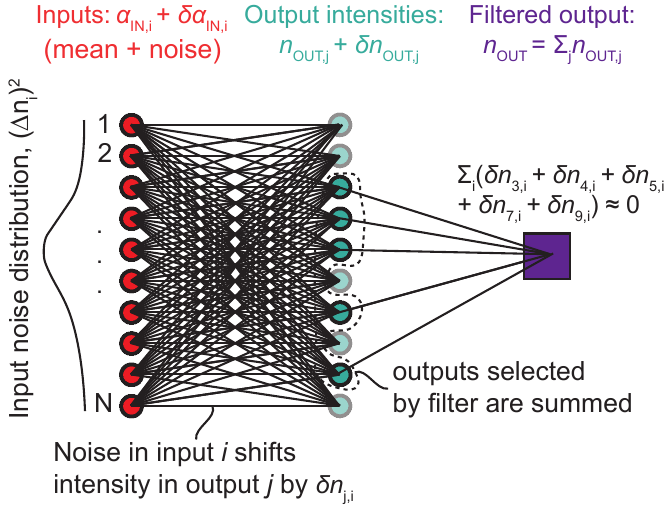}
     \caption{\textbf{Noise immunity due to multimode quantum correlations.} The noise in any output quantity such as the number of photons of a filtered output $n_{\text{OUT}}$, of a nonlinear interaction depends on both the magnitude of the noise (black curve) of the different input modes (red circles), and the sensitivity of the output to changes in the inputs. For an output that depends on multiple output modes, such as the sum of intensities in several modes (opaque teal circles), it is possible for a fluctuation in a mode $i$ to lead to nearly zero shift in the \emph{sum} of intensities (purple square), even if the individual modes shift strongly. This effect is due to quantum correlations created by the nonlinearity. Further, because of this perfect cancellation, the output noise is independent of the input level of noise. }
     \label{fig:fig4}
 \end{figure}

\begin{figure*}[t]
     \centering
     \includegraphics[width=0.9\textwidth]{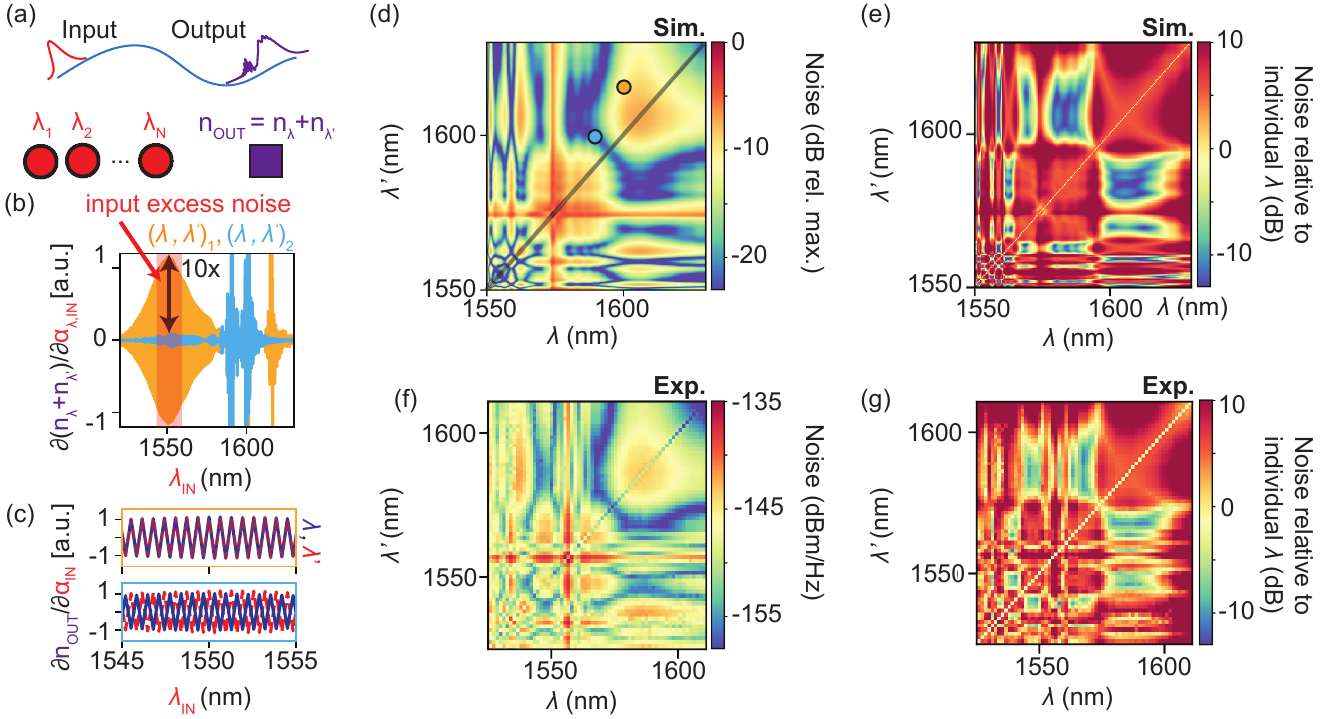}
     \caption{\textbf{Noise-immunity for pairs of colors induced by quantum correlations.} (a) Input-output formulation of ultrafast nonlinear dynamics of a femtosecond pulse in a fiber. The input modes are frequency components of the initial pulse. The output we measure is the sum of intensities in two wavelength channels of the output. (b) Simulated sensitivity of the total intensity in a pair of wavelengths, due to fluctuations of the input wavelength channels, as a function of the input wavelength $\lambda_{\text{IN}}$. The intensity of a typical pair of wavelengths $(\lambda,\lambda')_1$ (orange curve) displays a strong sensitivity to fluctuations in the wavelengths of the input pulse, shaded in red. The intensity in some special pairs $(\lambda,\lambda')_2$ (blue curve) displays strongly reduced sensitivity to changes in the input. (c) The reduced sensitivity occurs because a fluctuation at a given wavelength shifts the intensity in wavelengths $\lambda$ and $\lambda'$ oppositely for most fluctuation channels. Enhanced sensitivity comes from the same fluctuation of the input shifting the intensity in wavelengths $\lambda$ and $\lambda'$ in the same direction.  (d) Theoretically predicted variance in sum of intensities as a function of wavelengths $\lambda,\lambda'$. The blue and orange dots indicate the same wavelength pairs considered in (b). (e) Noise relative to the less-noisy of the two individual wavelengths. The signature of the cancellation in (b) is a much reduced noise (blue regions in (d)) and a much lower noise than the individual intensities (blue regions in (e). (f) Experimentally measured noise in the sum of intensities of different intensities. (g) Noise compared to the less noisy of the noise of the individual measured wavelengths.}
     \label{fig:fig5}
\end{figure*}

Our results are very likely to extend to a range of nonlinear systems. In fact, we expect that in a wide variety of multimode nonlinear systems, there are hidden low-noise states with quantum levels of noise, which only become accessible when filtering the light in the right basis. Similarly to the case of ``temporally multimode'' systems considered here, these low-noise states should emerge almost independently of noise in the input, enabling the development of a new generation of light sources which operate at the highest powers while also featuring genuinely quantum noise and correlations. 

We briefly discuss guidelines for finding these effects in other nonlinear systems. Finding the minimum noise corresponds to finding the filter function $T$ for which $(\Delta n)^2 = T\cdot(CT)$ is minimized, where $C$ is the intensity correlation between output modes. $C$ can either be measured (it is equivalent to the information in Figs. 5d,e) or predicted using Eq. (1). Using this, one can readily optimize for the best filter. Due to the intricate correlations created by nonlinearity, the filters can be complex (for example, having gaps like in Fig. 3). In many cases, the resulting output is adequate for many applications (e.g., if one is using the laser to pump a system with broadband response). While the question of how to realize the cleanest output with the highest power and lowest noise is an exciting direction for future work, we point out one possibility. By coherently shaping the input (e.g., controlling the intensity and phase of different inputs), one can tune the output correlations, allowing tunability in the intensity and noise of the output. 

Intense quantum light sources should facilitate the generation of quantum states of electrons, phonons, and other excitations in strongly driven materials. A number of effects associated with high-power lasers could be explored for this purpose, including topological solitons \cite{lumer2013self, mukherjee2020observation, jurgensen2023quantized}, self-similar systems \cite{fermann2000self, dudley2007self}, and spatially multimode systems \cite{wright2015controllable, wright2017spatiotemporal,wu2019thermodynamic,wright2020mechanisms, wright2022physics,pourbeyram2022direct, marques2023observation}. Even more broadly, there is a great range of effects which are being investigated, whose ``quantum optical nature'' either has not been touched at all, or contains many basic open questions. Examples of such systems and effects include complex parametric oscillators \cite{mcmahon2016fully, leefmans2022topological, roques2023biasing}, integrated squeezed light sources \cite{nehra2022few, guidry2023multimode, ng2023quantum, rivera2023creating, sloan2023driven},  disordered systems \cite{lahini2008anderson, lib2022quantum}, soliton microcombs \cite{kippenberg2018dissipative, guidry2022quantum, kues2019quantum}, nonlinear parity-time symmetric systems \cite{konotop2016nonlinear,xia2021nonlinear}, topological lasers \cite{harari2018topological, bandres2018topological}, topological spatial solitons \cite{mukherjee2020observation,jurgensen2021quantized,jurgensen2023quantized}, and high-harmonic generation \cite{gorlach2020quantum, lewenstein2021generation}. 

\textit{Acknowledgements.} We acknowledge useful discussions with Erich Ippen, Logan Wright, Ryotatsu Yanagimoto, Tatsuhiro Onodera, and Frank Wise. N.R. acknowledges the
support of a Junior Fellowship from the Harvard Society of
Fellows as well as funding from the School of Applied and Engineering Physics at Cornell University. Y. S. acknowledges support from the Swiss National Science Foundation (SNSF) through the Early Postdoc Mobility Fellowship No. P2EZP2-188091. J.S. acknowledges previous support
of a Mathworks Fellowship, as well as previous support from
a National Defense Science and Engineering Graduate (NDSEG) Fellowship (F-1730184536). This material is based upon work also supported in
part by the U. S. Army Research Office through the Institute for
Soldier Nanotechnologies at MIT, under Collaborative Agreement Number W911NF-23-2-0121. We also acknowledge support of
Parviz Tayebati.

\bibliographystyle{unsrt}
\bibliography{qnoise.bib}

\end{document}